\documentclass[aps,prl,onecolumn,groupedaddress]{revtex4}
\usepackage[]{graphicx}
\usepackage{amsmath}
\usepackage{amssymb}
\begin{document}

%\DOIsuffix{theDOIsuffix}
%%
%% issueinfo for header and copyright line
%\Volume{51}
%\Issue{1}
%\Month{01}
%\Year{2003}
%%
%%    First and last pagenumber of the article. If the option
%%    'autolastpage' is set (default) the second argument may be left empty.
%\pagespan{3}{}
%%
%%    Dates will be filled in by the publisher. The 'reviseddate' and
%%    'dateposted' (Published online) entry may be left empty.
\date{18 May 2006}
%\Reviseddate{}
%\Accepteddate{}
%\Dateposted{}
%%
%\keywords{quantum information, microtrap, atom chip, microwave potential, single atom detection, fiber cavity, optical lattice, Mott insulator.\\}

%\subjclass[pacs]{}

%% \pretitle{Editor's Choice}

\title{Quantum Information Processing in Optical Lattices
  and Magnetic Microtraps}

%% Please do not enter footnotes or \inst{}-notes into the optional
%% argument of the author command. The optional argument will go into
%% the header.  If there is only one address the marker \inst{x} may be
%% omitted.

%% Information for the first author.
\author{Philipp Treutlein\footnote{Corresponding author. E-mail: philipp.treutlein@physik.lmu.de, Phone: +49\,89\,2180-3937, Fax: +49\,89\,2180-3938}}

\author{Tilo Steinmetz}\altaffiliation[Present address: ]{Laboratoire Kastler Brossel de l'E.N.S, 24 Rue Lhomond, 75231 Paris Cedex 05, France}
\author{Yves Colombe}\altaffiliation[Present address: ]{Laboratoire Kastler Brossel de l'E.N.S, 24 Rue Lhomond, 75231 Paris Cedex 05, France}
\author{Benjamin Lev}\altaffiliation[Present address: ]{Dept. of Physics UCB/JILA, Boulder, CO 80309-0440, U.S.A.}
\author{Peter Hommelhoff}\altaffiliation[Present address: ]{Varian Physics Building, Stanford University, Stanford, CA 94305, U.S.A.}

\author{Jakob Reichel}\altaffiliation[Present address: ]{Laboratoire Kastler Brossel de l'E.N.S, 24 Rue Lhomond, 75231 Paris Cedex 05, France}

\author{Markus Greiner}\altaffiliation[Present address: ]{Harvard University, Department of Physics, % 17 Oxford Street,
Cambridge, MA 02138, U.S.A.}

\author{Olaf Mandel}\altaffiliation[Present address: ]{Varian Physics Building, Stanford University, Stanford, CA 94305, U.S.A.}

\author{Arthur Widera}\altaffiliation[Present address: ]{Institut f{\"u}r Physik, Johannes Gutenberg-Universit{\"a}t, % Staudingerweg 7,
55099 Mainz, Germany}
\author{Tim Rom}\altaffiliation[Present address: ]{Institut f{\"u}r Physik, Johannes Gutenberg-Universit{\"a}t, % Staudingerweg 7,
55099 Mainz, Germany}
\author{Immanuel Bloch}\altaffiliation[Present address: ]{Institut f{\"u}r Physik, Johannes Gutenberg-Universit{\"a}t, % Staudingerweg 7,
55099 Mainz, Germany}

\author{Theodor W. H{\"a}nsch}

\affiliation{Max-Planck-Institut f{\"u}r Quantenoptik und Sektion Physik der
Ludwig-Maximilians-Universit{\"a}t, Schellingstr. 4, 80799 M{\"u}nchen, Germany}

%%    \dedicatory{This is a dedicatory.}
\begin{abstract}

We review our experiments on quantum information processing with neutral atoms in optical lattices and magnetic microtraps.

Atoms in an optical lattice in the Mott insulator regime serve as a large qubit register. A spin-dependent lattice is used to split and delocalize the atomic wave functions in a controlled and coherent way over a defined number of lattice sites. This is used to experimentally demonstrate a massively parallel quantum gate array, which allows the creation of a highly entangled many-body cluster state through coherent collisions between atoms on neighbouring lattice sites.

In magnetic microtraps on an atom chip, we demonstrate coherent manipulation of atomic qubit states and measure coherence lifetimes exceeding one second at micron-distance from the chip surface. We show that microwave near-fields on the chip can be used to create state-dependent potentials for the implementation of a quantum controlled phase gate with these robust qubit states. For single atom detection and preparation, we have developed high finesse fiber Fabry-Perot cavities and integrated them on the atom chip. We present an experiment in which we detected a very small number of cold atoms magnetically trapped in the cavity using the atom chip.

\end{abstract}

\maketitle

%% If there is not enough space inside the running head
%% for all authors including the title you may provide
%% the leftmark in one of the following three forms:

%% \renewcommand{\leftmark}
%% {F. Author: A short title}

%% \renewcommand{\leftmark}
%% {F. Author and S. Author: A short title}

 \renewcommand{\leftmark}
 {P. Treutlein et al.: Quantum Information Processing in Optical Lattices
  and Magnetic Microtraps}

\tableofcontents  % Produces the table of contents.

\section{Introduction}
Neutral atoms present two essential advantages for quantum information
processing (QIP). They are relatively weakly coupled to the
environment, so that decoherence can be controlled better than in most
other systems. Furthermore, complete control of all quantum-mechanical
degrees of freedom is already a reality, and is used in experiments
with great success, most notably in Bose-Einstein condensation.

Theoretical approaches have been developed to use atoms in
well-defined states of controllable potentials for creating
many-particle entanglement, and qubit operations in particular.
Experimentally realizing these proposals is a major challenge and
requires new ideas to overcome the subtle problems occuring in real
atomic systems. Requirements on stability and control of environmental
conditions, such as electric and magnetic stray fields, are equally
demanding.

In the theoretical investigations already, two experimental systems
emerged as particularly promising embodiments for neutral-atom QIP.
{\it Optical lattices} allow for a large number of qubits due to their
three-dimensional, periodic structure. In {\it magnetic microtraps
  (atom chips)}, complex potentials can be realized, and lithographic
fabrication techniques enable scalability and modularity in analogy
with microelectronics. In the following, we review experimental
progress achieved in our group with both systems.

\section{Optical lattices}

\subsection{Preparation of a qubit register}
\begin{figure}[h]
\begin{center}
\includegraphics[scale=0.5]{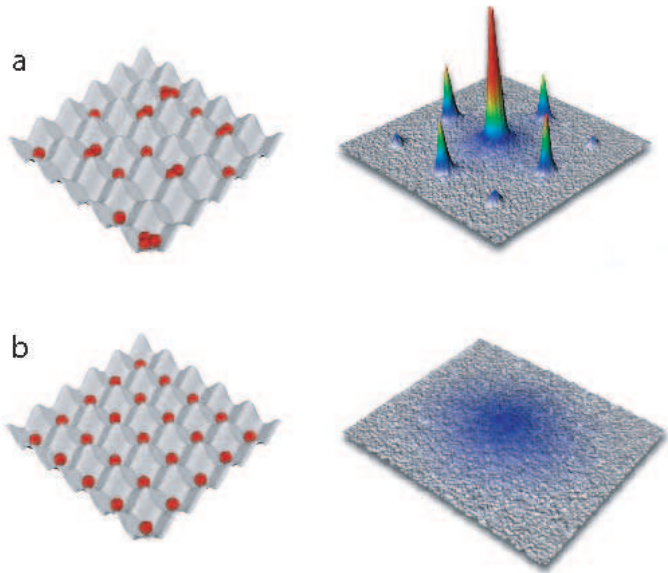}
\caption{(a) In the superfluid state of a Bose-Einstein condensate, the underlying atoms can be described as a giant macroscopic matter wave. When such a condensate is released from the periodic potential a multiple matter wave interference pattern is formed due to the phase coherence between the atomic wavefunctions on different lattice sites. In this case the phase of the macroscopic matter wave is well defined. However, the number of atoms at each lattice site fluctuates. (b) In the limit of a Mott insulating state, each lattice site is filled with a fixed number of atoms but the phase of the matter wave field remains uncertain. As a result, no matter wave interference pattern can be seen when the quantum gases are released from the lattice potential. }
\label{latticefig:0}
\end{center}
\end{figure}

Starting point for the preparation of the neutral atom qubit register is an atomic Bose-Einstein condensate. This is placed in an artificial crystal of light - a so called optical lattice - which is formed
by standing wave laser fields along all three space dimensions. By continuously increasing the lattice depth of the optical potentials, one can drive the system through a quantum phase transition from a superfluid to a Mott insulator \cite{Jaksch98,Greiner02a}, where a defined number of atoms is placed on each lattice site (see Fig.÷\ref{latticefig:0}). By controlling the initial total
number of atoms and the confinement parameters of the lattice trap, it is possible to have a large connected region to be populated by single atoms on each lattice site. On each of these sites, the atoms occupy the ground state of the trapping potential and their internal state
is initialized to a defined state as well.

\subsection{A quantum conveyer belt for neutral atoms}
So far the optical potentials used for optical lattices with
Bose-Einstein condensates have been mostly independent of the
internal ground state of the atom. However, it has been suggested
that by using spin-dependent periodic potentials one could bring
atoms on different lattice sites into contact and thereby realize
fundamental quantum gates
\cite{Brennen99,Briegel00,Raussendorf01,Brennen02}, create large
scale entanglement \cite{Jaksch99,Briegel01}, excite spin waves
\cite{Sorensen99}, study quantum random walks \cite{Duer02} or form
a universal quantum simulator to simulate fundamental complex
condensed matter physics hamiltonians \cite{Jane02}. Here we show
how the wave packet of an atom that is initially localized to a
single lattice site can be split and delocalized in a controlled and
coherent way over a defined number of lattice sites.

\begin{figure}
\begin{center}
\includegraphics[scale=0.7]{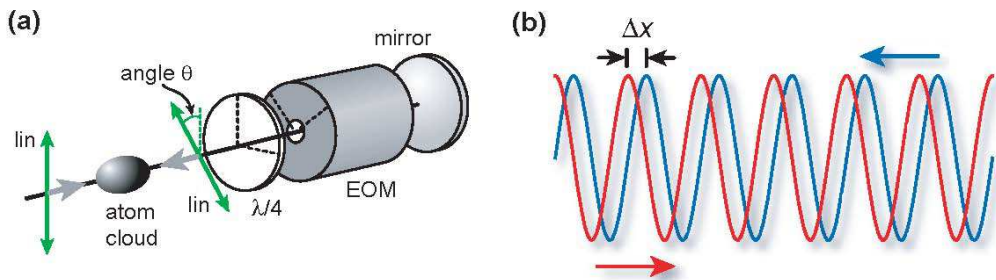}
\caption{\label{fig:expsetup} {\bf (a)} Schematic experimental
setup. A one dimensional optical standing wave laser field is formed
by two counterpropagating laser beams with linear polarizations. The
polarization angle of the returning laser beam can be adjusted
through an electro-optical modulator. The dashed lines indicate the
principal axes of the wave plate and the EOM. {\bf (b)} By
increasing the polarization angle $\theta$, one can shift the two
resulting $\sigma^+$ (blue) and $\sigma^-$ (red) polarized standing
waves relative to each other.}\label{fig:expsetupspindep}
\end{center}
\end{figure}

In order to realize a spin dependent transport for neutral atoms in
optical lattices, a standing wave configuration formed by two
counterpropagating laser beams with linear polarization vectors
enclosing an angle $\theta$ has been proposed
\cite{Brennen99,Jaksch99}. Such a standing wave light field can be
decomposed into a superposition of a $\sigma^+$ and $\sigma^-$
polarized standing wave laser field, giving rise to lattice
potentials $V_+(x,\theta)=V_0 \cos^2(kx+\theta/2)$ and
$V_-(x,\theta)=V_0 \cos^2(kx-\theta/2)$. By changing the
polarization angle $\theta$, one can control the separation
between the two potentials $\Delta x =\theta/180^\circ \cdot
\lambda_x/2$ (see Fig.~\ref{fig:expsetupspindep}b). When increasing
$\theta$, both potentials shift in opposite directions and overlap
again when $\theta=n\cdot180^\circ$, with $n$ being an integer. For
a spin-dependent transfer, two internal spin states of the atom
should be used, where one spin state dominantly experiences the
$V_+(x,\theta)$ dipole potential and the other spin state mainly
experiences the $V_-(x,\theta)$ potential. Such a
situation can be realized in rubidium by tuning the wavelength of
the optical lattice laser to a value of $\lambda_x=785$\,nm between
the fine structure splitting of the rubidium D1 and D2 transition.
If an atom is now first placed in a coherent superposition of both
internal states $1/\sqrt{2}(|0\rangle+i|1\rangle)$ and the
polarization angle $\theta$ is continuously increased, the spatial
wave packet of the atom is split with both components moving in
opposite directions.

\begin{figure}[h]
\begin{center}
\begin{tabular}{ll}
(i) & (ii) \\
\includegraphics[scale=0.4]{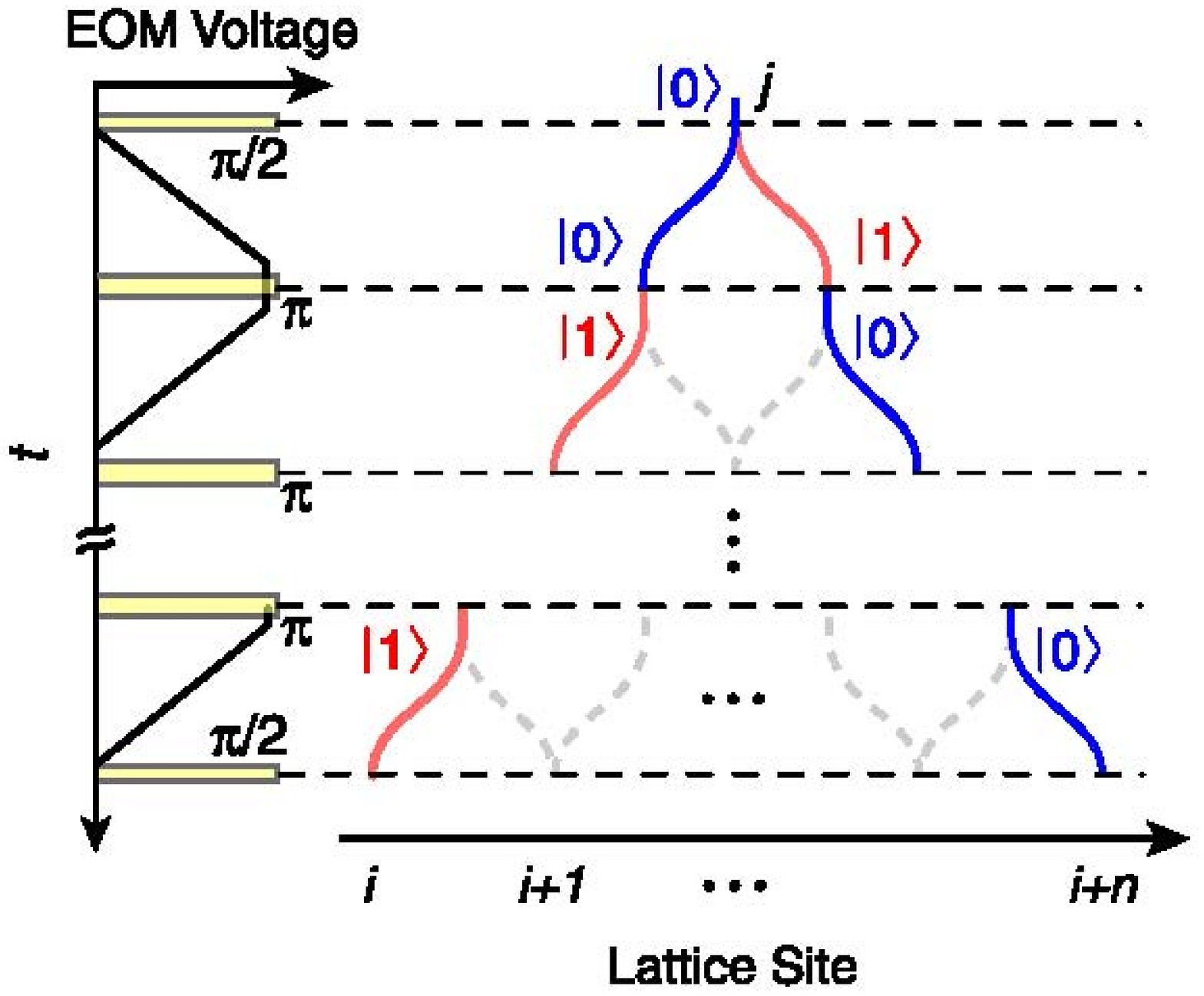} & \includegraphics[scale=0.4]{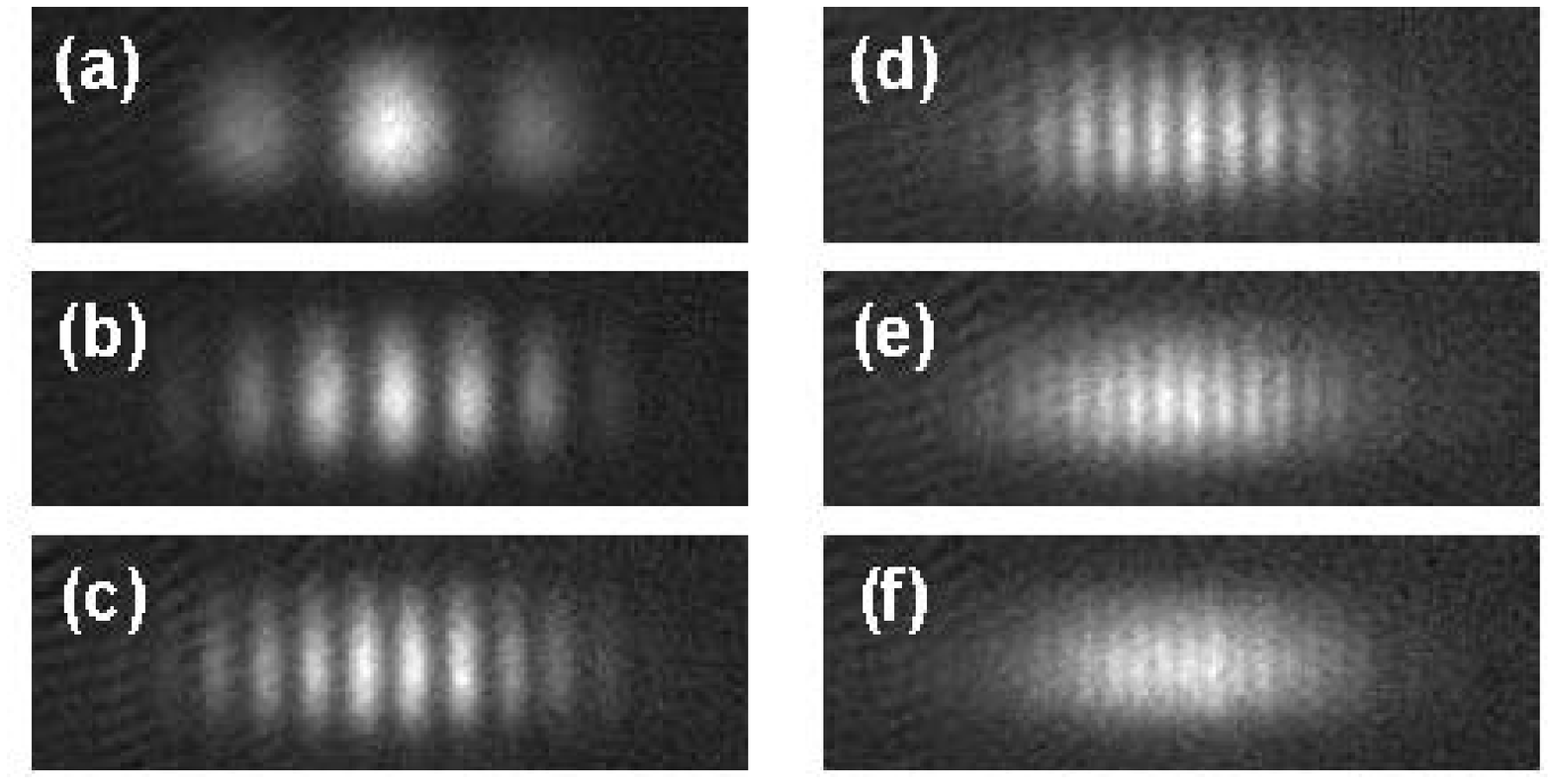}
\end{tabular}
\caption{(i) Schematic sequence used for the quantum conveyer belt. A single atom on lattice site j can be transported over an arbitrary number of lattice sites depending on its spin state (marked as blue and red curves). (ii) This has allowed us to split the wave function of the atom in a coherent way, such that a single atom simultaneously moves to the left and to the right. The coherence of the split wave-packets has been demonstrated in an interference experiment. For larger distances between the split wave-functions, the period of the interference pattern decreases. }
\label{latticefig:1}
\end{center}
\end{figure}

With such a quantum conveyer belt, atoms have been moved over a defined number of lattice sites. In the experiment a coherent transport of the atoms over a distance of up to 7 lattice sites has been demostrated \cite{Mandel03a} (see Fig.\,\ref{latticefig:1}).

\subsection{Controlled collisions}
In order to realize a controlled interaction between the particles
on different lattice sites in a 3D Mott insulating quantum register,
the above spin dependent transport sequence can be used. This leads
to collisions between neighbouring atoms and can be described
through an ensemble of quantum gates acting in parallel
\cite{Jaksch99,Briegel00}. Alternatively, these quantum gates can be
described as a controllable quantum Ising interaction
\cite{Briegel01}:

\begin{equation}
H_\mathrm{int} \propto g(t) \sum_j \frac{1+\sigma_z^{(j)}}{2}
\frac{1-\sigma_z^{(j+1)}}{2}
\end{equation}

\begin{figure}
\begin{center}
\includegraphics[scale=0.7]{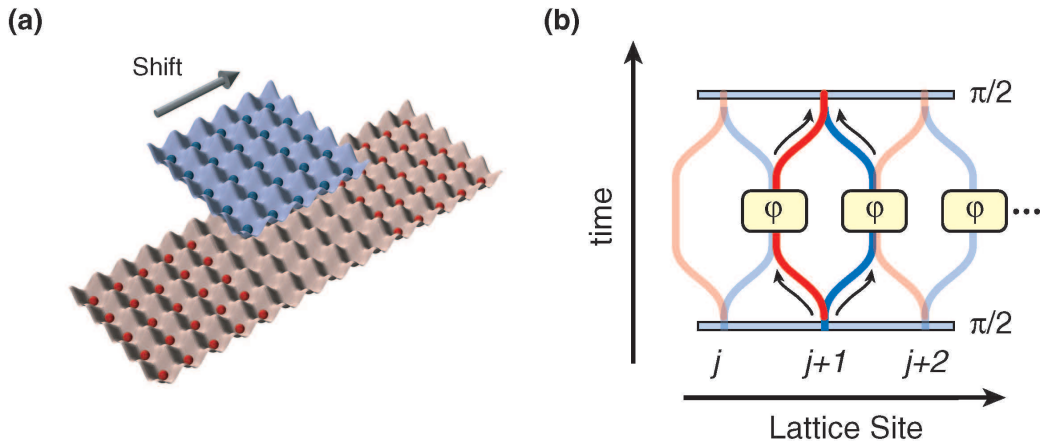}
\caption{\label{fig:collisions_latticeshift}{\bf (a)} Controlled
interactions between atoms on different lattice sites can be
realized with the help of spin-dependent lattice potentials. In such
spin dependent potentials, atoms in a, let us say, blue internal
state experience a different lattice potential than atoms in a red
internal state. These lattices can be moved relative to each other
such that two initially separated atoms can be brought into
controlled contact with each other. {\bf(b)} This can be extendended
to form a massively parallel quantum gate array. Consider a string
of atoms on different lattice sites. First the atoms are placed in a
coherent superposition of the two internal states (red and blue).
Then spin dependent potentials are used to split each atom such that
it simultaneously moves to the right and to the left and is brought
into contact with the neighbouring atoms. There both atoms interact
and a controlled phase shift $\varphi$ is introduced. After such a
controlled collision the atoms are again moved back to their
original lattice sites.}\label{fig:controlledcollisions_schematic}
\end{center}
\end{figure}

Here $g(t)$ denotes the time dependent coupling constant and $\sigma_z^{(j)}$  is
the Pauli spin operator acting on an atom at the $j^{th}$ lattice
site. For an interaction phase of
$\varphi=2\pi\times\int_0^{t_\mathrm{hold}} g(t) \,dt/h=(2n+1)\pi$ one
obtains a maximally entangled cluster state, whereas for $\varphi
=2n \pi$ one obtains a disentangled state \cite{Briegel01}. Here
$t_\mathrm{hold}$ denotes the time for which the atoms are held together at
a common site, $h$ is Planck's constant and $n$ is an integer. Let
us point out that the creation of such highly entangled states can
be achieved in a single lattice shift operational sequence described
above and depicted in Fig.~\ref{fig:controlledcollisions_schematic},
independent of the number of atoms to be entangled
\cite{Jaksch99,Briegel01}.

A $\pi/2$ pulse allows us to place the atom in a coherent
superposition of the two states $|0\rangle \equiv |F=1,m_F=-1\rangle$ and $|1\rangle \equiv |F=2,m_F=-2\rangle$ within a
time of 6\,$\mu$s. After creating such a coherent superposition, we
use a spin-dependent transfer to split and move the spatial wave
function of the atom over half a lattice spacing in two opposite
directions depending on its internal state (see
Fig.~\ref{fig:controlledcollisions_schematic}). Atoms on
neighbouring sites interact for a variable amount of time $t_\mathrm{hold}$
that leads to a controlled conditional phase shift of the
corresponding many body state. After half of the hold time, a
microwave $\pi$  pulse is furthermore applied. This spin-echo type
pulse is mainly used to cancel unwanted single particle phase shifts
e.g. due to inhomogeneities in the trapping potentials. It does not,
however, affect the non-trivial and crucial collisional phase shift
due to the interactions between the atoms. After such a controlled
collision, the atoms are moved back to their original site. Then a
final $\pi/2$ microwave pulse with variable phase is applied and the
atom number in state $|1\rangle$ relative to the total atom number is recorded \cite{Mandel03b}.

\begin{figure}[h]
\begin{center}
\includegraphics[scale=0.35]{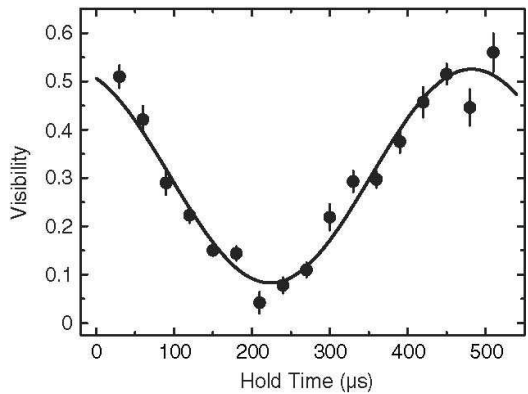}
\caption{Visibility of Ramsey fringes vs. hold times on neighbouring
lattice sites for the experimental sequence of
Fig.~\ref{fig:controlledcollisions_schematic}. The solid line is a
sinusoidal fit to the data including an offset and a finite
amplitude. Such a sinusoidal behaviour of the visibility vs. the
collisional phase shift (determined by the hold time $t_\mathrm{hold}$) is
expected for a Mott insulating state with an occupancy of n=1 atom
per lattice site.}\label{fig:entanglementoscillations}
\end{center}
\end{figure}

For short hold times, where no significant collisional phase shift
is acquired, a Ramsey fringe with a high visibility of approx. 50\%
is recorded (see Fig.~\ref{fig:entanglementoscillations}). For
longer hold times, we notice a strong reduction in the visibility of
the Ramsey fringe, with an almost vanishing visibility of approx.
5\% for a hold time of 210\,$\mu$s. This hold time corresponds to an
acquired collisional phase shift of $\varphi = \pi$ for which we
expect a minimum visibility if the system is becoming entangled. For
a two-particle system this can be understood by observing the
resulting Bell state:

\begin{equation}
1/\sqrt{2} \left ( |0\rangle_j |+\rangle_{j+1}^\alpha + |1\rangle_j
|-\rangle_{j+1}^\alpha \right ),
\end{equation}

after the final $\pi/2$ pulse of the Ramsey sequence has been
applied to the atoms. Here $|+\rangle_{j+1}^\alpha$ and
$|-\rangle_{j+1}^\alpha$ represent two orthogonal superposition
states of $|0\rangle$ and $|1\rangle$ for which $|\langle 1
|+\rangle^\alpha|^2+|\langle 1 |-\rangle^\alpha|^2=0.5$. A
measurement of atoms in state $|1\rangle$ therefore becomes
independent of the phase corresponding to a vanishing Ramsey fringe.
This indicates that no single particle operation can place all atoms
in either spin-state when a maximally entangled state has been
created. The disappearance of the Ramsey fringe has been shown to
occur not only for a two-particle system, but is a general feature
for an arbitrary $N$-particle array of atoms that have been highly
entangled with the above experimental sequence \cite{Briegel00}. For
longer hold times however, the visibility of the Ramsey fringe
increases again reaching a maximum of 55\% for a hold time of
450\,$\mu$s (see Fig.~\ref{fig:entanglementoscillations}). Here the
system becomes disentangled again, as the collisional phase shift is
close to $\varphi=2\pi$ and the Ramsey fringe is restored with
maximum visibility. The timescale of the observed collisional phase
evolution is in good agreement with the measurements on the Mott
insulator transition of the previous section and ab-initio
calculations of the onsite matrix element $U$ \cite{Jaksch98,Greiner02a}.

\section{Magnetic microtraps}
Atom chips \cite{Reichel02,FolmanReview02} combine many important features of a scalable
architecture for quantum information processing \cite{diVincenzo00}: The long
coherence lifetimes of qubits based on hyperfine states of neutral
atoms \cite{Treutlein04}, accurate control of the coherent evolution of the atoms in
tailored micropotentials \cite{Hommelhoff05,Schumm05}, and scalability of the technology through
microfabrication \cite{Lev03,Groth04} -- which allows the integration of many qubits
in parallel on the same device while maintaining individual
addressability. Furthermore, atom chips offer the exciting perspective of creating interfaces between the atomic qubits and other QIP systems integrated on the same chip, such as photons in optical fiber cavities or solid-state QIP systems located on the chip surface \cite{Sorensen04}. However, the experimental demonstration of a fundamental
two-qubit quantum gate on an atom chip is an important milestone
which still has to be reached.

In \cite{Calarco00}, a first theoretical proposal for a quantum gate on an atom chip was put forward. In this proposal, the gate operation relies on collisional interactions between two atoms in a state-selective potential on the chip. The experimental challenge of implementing such a gate can be divided into several steps:
%The experimental challenge of implementing a quantum gate on an atom chip can be divided into several steps:

\begin{enumerate}
\item
A qubit state pair has to be identified which can be manipulated with electromagnetic fields on the atom chip, but still allows for long coherence lifetimes in a realistic experimental situation. In particular, attention has to be paid to decoherence and loss mechanisms induced by the chip surface, which is typically at a distance of only few microns from the atoms.
%High fidelity single qubit rotations between the qubit states have to be demonstrated.

\item
The gate proposed in \cite{Calarco00} requires potentials which affect the two qubit states differently in order to achieve conditional logical operations between two atoms. A method to create the required potentials on a chip has to be developed.

%A configuration of potentials on the chip has to be found in which conditional logic between two atomic qubits can be implemented.
%A quantum gate on an atom chip can be implemented via collisional interactions between two atoms in a state-selective potential on the atom chip \cite{Calarco00}. Therefore, a method to create state-selective potentials for the qubit state pair has to be developed
%A method to created potentials which affect the two qubit states differently has to be developed.

\item
While Bose-Einstein condensates and thermal ensembles of atoms are routinely manipulated and detected on atom chips, the existing proposals for quantum information processing on atom chips rely on coherent control over single atoms. As a first step towards single atom operation, a single atom detector has to be developed which can be integrated on the atom chip.

\item
With a single atom detector available, a method for the deterministic preparation of single neutral atoms in the motional ground state of chip traps with very low occupation probability of excited states has to be found.

\end{enumerate}

In the following, we develop a scenario in which these challenges can be met with atom chips and discuss our experiments towards its realization.
%In particular, we discuss possible solutions to the challenges 1-3, which we have studied in our experiments. We conclude with a few remarks on challenge 4.

\subsection{Qubit states on the atom chip}

Two conflicting requirements have to be met by the qubit states $\{|0\rangle, |1\rangle \}$ chosen for QIP on an atom chip. On the one hand, both qubit states have to couple to electromagnetic fields which are used for trapping and manipulating the atoms. In all experiments performed so far, at least a part of the trapping potential is provided by static magnetic fields generated by wires or permanent magnet structures on the atom chip. It is therefore desirable that both $|0\rangle$ and $|1\rangle$ are magnetically trappable states.
On the other hand, gate operations with high fidelity are only possible if the coherence lifetimes of the superposition states $\alpha |0\rangle + \beta |1\rangle$, $(|\alpha |^2 + |\beta |^2 = 1)$ are sufficiently long. Long coherence lifetimes are possible if qubit basis states are chosen whose energy difference $h\nu_{10} = E_{|1\rangle}-E_{|0\rangle}$ is robust against noise in realistic experimental situations. In particular, technical fluctuations of magnetic fields are notorious for limiting the coherence lifetime of magnetic-field sensitive qubit states of atoms or ions to a few milliseconds \cite{SchmidtKaler03}. On atom chips, magnetic near-field noise due to thermally excited currents in the chip wires is an additional fundamental source of decoherence for magnetic field sensitive qubit states \cite{Henkel03}. To achieve long coherence lifetimes on atom chips, it is therefore highly desirable to choose a pair of qubit basis states whose energy difference is insensitive to magnetic field fluctuations.

We choose the $|F=1,m_F=-1\rangle \equiv |0\rangle$ and $|F=2,m_F=+1\rangle \equiv |1\rangle$ hyperfine levels of the $5S_{1/2}$ ground state of $^{87}$Rb atoms as qubit basis states. The magnetic moments and the corresponding static Zeeman shifts of the two states are approximately equal, leading to a strong common mode suppression of magnetic field induced decoherence. Furthermore, both states experience nearly identical trapping potentials in magnetic traps, thereby avoiding undesired entanglement between internal and external degrees of freedom of the atoms.
At a magnetic field of $B_0 \sim 3.23$\,G, both states experience the same first-order Zeeman shift and the remaining magnetic field dependence of the transition frequency $\nu_{10}$ is minimized \cite{Harber02}. In all of our experiments, we therefore adjust the
field in the center of the trap to $B_0$.

We have studied the coherence properties of the state pair $\{|0\rangle,|1\rangle\}$ on an atom chip in a series of experiments \cite{Treutlein04}, which we summarize in the following.

The coherence measurements are performed with an ultracold ensemble of atoms, which is prepared in a multi-step sequence involving loading of the microtrap from a mirror-MOT, compression of the trap and evaporative cooling \cite{Haensel01a}. By the end of this sequence, a thermal atomic ensemble of typically $N_\mathrm{at}=1.5 \times 10^4$ atoms in state $|0\rangle$ at a temperature of $0.6\,\mu$K is trapped in a Ioffe-type microtrap. By modulating the currents and offset magnetic fields used to create this trap, the atoms can be placed at distances $d=0-130\,\mu$m from the chip surface with only small changes in the shape of the magnetic potential. It is advantageous to perform the coherence measurements with a small thermal ensemble instead of a Bose-Einstein condensate, since the higher atomic densities in the condensate would lead to a stronger inhomogeneous collisional broadening of the qubit transition \cite{Harber02}.

\begin{figure}[]
\begin{center}
\includegraphics[scale=0.75]{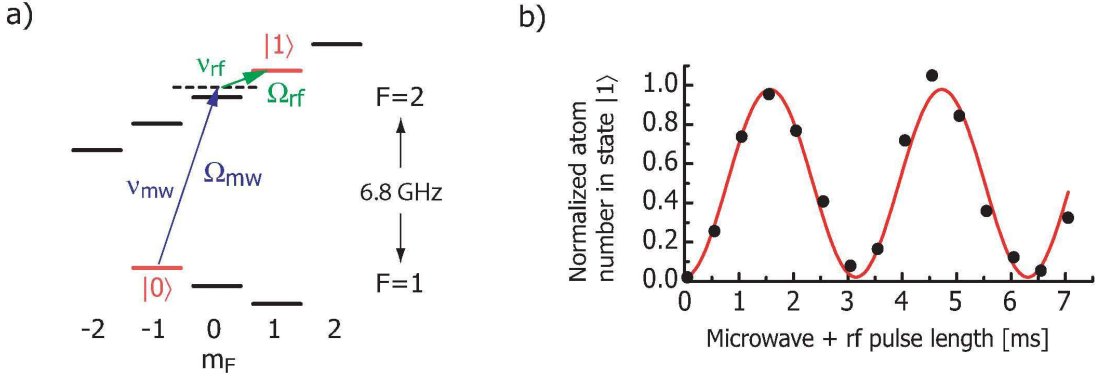}
\end{center}
\caption{\label{fig:Rabi} Single qubit rotations between the states $|0\rangle$ and $|1\rangle$. (a) Ground state hyperfine structure of $^{87}$Rb in a weak magnetic field. The first order Zeeman shift of the states $|0\rangle$ and $|1\rangle$ is approximately identical. The two-photon transition $|0\rangle \leftrightarrow |1\rangle$ is driven by a microwave $\nu_\mathrm{mw}$ and a radio-frequency $\nu_\mathrm{rf}$. $\Omega_\mathrm{mw}$ and $\Omega_\mathrm{rf}$ are the single-photon Rabi frequencies of the microwave and rf transition, respectively. (b) Two-photon Rabi oscillations recorded as a function of the microwave and rf pulse length. The two-photon Rabi frequency is $\Omega_\mathrm{2ph}/2\pi = 0.32$\,kHz.}
\end{figure}

Single-qubit rotations are implemented by coupling the states $|0\rangle$ and $|1\rangle$ through a two-photon microwave-rf transition as shown in Fig.~\ref{fig:Rabi}a. The microwave frequency $\nu_\textrm{mw}$ is detuned by $\delta/2\pi=1.2$\,MHz above the $|F=2,m_F=0\rangle$ intermediate state and radiated from a sawed-off waveguide outside the vacuum chamber. The radio frequency $\nu_\textrm{rf}$ is either applied to an external coil or to a wire on the chip. $\nu_\textrm{mw}$ and $\nu_\textrm{rf}$ are phase locked to a 10\,MHz reference from an ultrastable quartz oscillator.  $\Omega_\mathrm{mw}$ and $\Omega_\mathrm{rf}$ are the single-photon Rabi frequencies of the microwave and rf transition, respectively. By applying the two-photon drive for a variable time and detecting the number of atoms $N_1$ transferred from $|0\rangle$ to $|1\rangle$, we observe Rabi oscillations with a resonant two-photon Rabi frequency of $\Omega_\mathrm{2ph}/2\pi = 0.32$\,kHz, see Fig.~\ref{fig:Rabi}b. The maximum transition probability, corresponding to a $\pi$-pulse, is $N_1/N_\mathrm{at}=95\pm 5$\,\%.

The two-photon Rabi frequency is given by $\Omega_\mathrm{2ph} = \Omega_\mathrm{mw}\Omega_\mathrm{rf}/2\delta$ if $\Omega_\mathrm{mw}^2, \Omega_\mathrm{rf}^2 \ll \delta^2$ \cite{Gentile89}, with $\Omega_\mathrm{mw} \sim \Omega_\mathrm{rf} \sim 2\pi \times 25$\,kHz in our experiment. In the present experiment, the two-photon Rabi-frequency is limited by the available microwave power of typically a few watts. Instead of radiating the microwave and rf from antennas outside the vacuum chamber, they can be applied to the atoms much more efficiently by coupling the microwave and rf signal into wires designed as waveguiding structures on the chip. Consider a waveguide on a chip with a characteristic impedance of $Z_c=50\,\Omega$ carrying a microwave signal of $P=1$\,mW, corresponding to a microwave current of $I_\mathrm{mw} = \sqrt{2 P/Z_c} = 6.3$\,mA on the signal conductor. At a distance of $d=10\,\mu$m from the signal conductor, the microwave magnetic field amplitude is approximately $B_\mathrm{mw} \sim \mu_0 I_\mathrm{mw}/( 2 \pi d) = 1.3$\,G. The microwave induces a coupling with a single-photon Rabi frequency of the order of $\Omega_\mathrm{mw}/2\pi \sim \mu_B B_\mathrm{mw} / h = 1.8$\,MHz. This shows that it is advantageous to couple the atomic transitions with microwave and rf near fields instead of radiation from antennas.

%The microwave magnetic field amplitude at a distance of $d=10\,\mu$m from a thin wire carrying a microwave or rf signal of $P=1$\,mW at an impedance of $Z_c=50\,\Omega$ is given by $B_\mathrm{mw}= \mu_0 \sqrt{P/Z_c}\, /\, ( 2 \pi d) = 0.9$\,G. This field induces a coupling with single-photon Rabi frequency $\Omega_\mathrm{mw}/2\pi \simeq \mu_B B_\mathrm{mw} / h = 1.3$\,MHz.

To test for decoherence of the superposition states, we perform Ramsey spectroscopy by applying the following pulse sequence: The atoms in state $|0\rangle$ are held in the trap before a first $\pi/2$-pulse creates a coherent superposition of $|0\rangle$ and $|1\rangle$. After a time delay $T_R$, a second $\pi/2$-pulse is applied, and the resulting state is probed by detecting the number of atoms transferred to state $|1\rangle$. Ramsey fringes are recorded in the time domain by varying $T_R$ while keeping $\delta_R=\nu_\textrm{mw}+\nu_\textrm{rf} - \nu_{10}$ fixed ($\delta_R\ll \nu_{10}\simeq 6.8$\,GHz). Alternatively, Ramsey fringes are recorded in the frequency domain by scanning $\delta_R$ while $T_R$ remains constant. Loss of coherence of the superposition states can show up in different ways in the Ramsey signal. A spatial variation of $\nu_{10}$ across the atomic ensemble leads to a decay of the contrast of the Ramsey fringes, while temporal fluctuations of $\nu_{10}$ lead to increasing phase noise on the Ramsey oscillation as $T_R$ is increased.

\begin{figure}[]
\begin{center}
\includegraphics[scale=0.8]{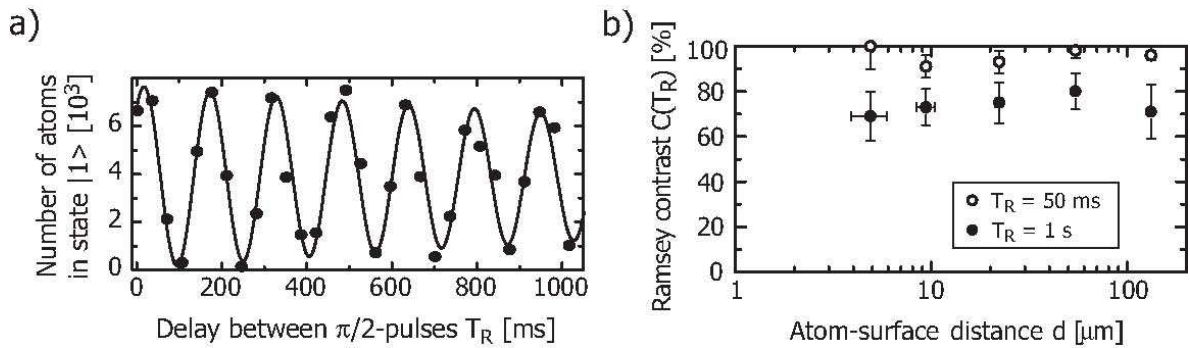}
\end{center}
\caption{\label{fig:Ramsey} Coherence lifetime measurements for the qubit state pair. (a) Ramsey spectroscopy of the $|0\rangle \leftrightarrow |1\rangle$ transition with atoms held at a distance $d=9\,\mu$m from the chip surface. An exponentially damped sine fit to the Ramsey fringes yields a $1/e$ coherence lifetime of $\tau_c=2.8\pm 1.6$\,s. Each data point corresponds to a single shot of the experiment. (b) Contrast $C(T_R)$ of the Ramsey fringes as a function of atom-surface distance $d$ for two values of the time delay $T_R$ between the $\pi/2$-pulses. For each data point, $C(T_R)=(N_\textrm{max}-N_\textrm{min})/(N_\textrm{max}+N_\textrm{min})$ was obtained from a sinusoidal fit to frequency-domain Ramsey fringes. $N_\textrm{max}$ ($N_\textrm{min}$) is the maximum (minimum) of the oscillation in $N_1$.
%The data points for $d=(5,9,22,54,132)\,\mu$m were measured with atomic ensembles of temperatures $T=(0.2,0.6,0.7,0.6,0.3)\mu$K and peak densities $n_0=(4,3,1,1,5)\times 10^{12}$\,cm$^{-3}$.
}
\end{figure}

Figure~\ref{fig:Ramsey}a shows Ramsey interference in the time domain. The number of atoms detected in state $|1\rangle$ oscillates at the frequency difference $\delta_R=6.4$\,Hz, while the interference contrast decays with a coherence lifetime of $\tau_c=2.8\pm 1.6$\,s. The measurement shown in Fig.~\ref{fig:Ramsey}a was performed at a distance $d=9\,\mu$m from the room-temperature chip surface. In \cite{Harber02}, similar coherence lifetimes are reported for the same state pair, but with atoms in a macroscopic magnetic trap, far away from any material objects. This suggests that atom-surface interactions indeed do not limit the coherence lifetime in our present experiment.

To further probe for surface effects, we study the decoherence of the Ramsey signal as a function of atom-surface distance. Atomic ensembles are prepared in traps at different distances $d$ from the surface. In each trap, we record Ramsey oscillations in the frequency domain for several values of $T_R$ and determine the contrast $C(T_R)$ of each oscillation. Figure~\ref{fig:Ramsey}b shows the result of these measurements for $T_R=50$\,ms and $T_R=1$\,s. Within the experimental error, the contrast does not show a dependence on atom-surface distance for $d=5-130\,\mu$m.  Additionally, we have compared the signal-to-noise ratio $S/N$ of the interference signal in the different traps. We typically observe $S/N=6$ for $T_R=1$\,s, where $S$ is the peak-to-peak amplitude of the sinusoidal fit to the Ramsey oscillation and $N$ is the standard deviation of the fit residuals over one oscillation period. $S/N$ is independent of $d$ within experimental error, indicating that the processes causing amplitude and phase fluctuations of the interference signal do not depend on atom-surface distance on this time scale. The observed noise on the Ramsey oscillation is mostly phase noise and can be attributed to ambient magnetic field fluctuations, which are independent of atom-surface distance.

%The observed decoherence is mainly due to a combination of the residual differential Zeeman shift and the density dependent collisional shift of $\nu_{10}$ across the ensemble \cite{Harber02}. Consequently, we observe a dependence of the coherence lifetime on the temperature $T$ and on the peak density $n_0$ of the ensemble. To avoid systematic errors, we have checked that there is no systematic variation of $T$ and $n_0$ in the measurement trap as $d$ is varied (see Fig.~\ref{fig:Ramsey}b). The observed noise on the Ramsey oscillation is mostly phase noise and can be attributed to ambient magnetic field fluctuations.

Our experiments show that the robust qubit state pair considered here can be manipulated on the atom chip with coherence lifetimes $\tau_c > 1$\,s at distances down to a few microns from the chip surface. In the proposal for a quantum controlled phase gate on an atom chip \cite{Calarco00}, a gate operation time of $\tau_g = 0.4$\,ms was estimated. Implementing this gate with our qubit state pair, $\tau_c/\tau_g \sim 10^3$ gate operations could be performed before decoherence from magnetic noise occurs.
In contrast, in the original proposal of \cite{Calarco00}, the qubit is encoded in two states with a magnetic-field sensitive energy difference. The magnetic field sensitivity is more than a factor of $10^3$ higher than for our state pair, so that expected coherence lifetimes would be comparable to the gate operation time.
%Compared to \cite{Calarco00}, the decoherence rate due to magnetic-field noise of our qubit state pair is suppressed by a factor of $10^{-6}$.

\subsection{State-dependent microwave potentials}

An implementation of the phase gate proposed in \cite{Calarco00} with our qubit state pair  requires potentials which affect the two qubit states differently. However, a combination of static magnetic and electric fields, as considered in \cite{Calarco00,Krueger03}, does not provide state-selective potentials for our state pair, whose magnetic moments and electrostatic polarizabilities are equal. Optical potentials created by focussed laser beams with a frequency close to the D1 or D2 transition of $^{87}$Rb are also impractical: if the detuning of the laser from the atomic resonance is much larger than the hyperfine splitting of the $^{87}$Rb ground state, the resulting optical potentials are again nearly identical for the states $|0\rangle$ and $|1\rangle$. If, on the other hand, a detuning comparable to the hyperfine splitting is used, a differential optical potential could be created, but problems with decoherence due to spontaneous scattering of photons would arise.

To generate the state-dependent potential for our qubit, we propose to use microwave potentials in addition to static magnetic potentials on the atom chip \cite{Treutlein04}. Microwave potentials arise due to the AC Zeeman effect (the magnetic analog of the AC Stark effect) induced by tailored microwave near-fields. In $^{87}$Rb, microwave potentials derive from magnetic dipole transitions with a frequency near $\omega_0/2\pi= 6.835$\,GHz between the $F=1$ and $F=2$ hyperfine manifolds of the ground state.  The magnetic component of a microwave field of frequency $\omega_\mathrm{mw}=\omega_0+\Delta$ couples the $|F=1,m_1\rangle$ to the $|F=2,m_2\rangle$ sublevels and leads to energy shifts that depend on $m_1$ and $m_2$. In a spatially varying microwave field, this results in a state-dependent potential landscape.

\begin{figure}[]
\begin{center}
\includegraphics[scale=0.7]{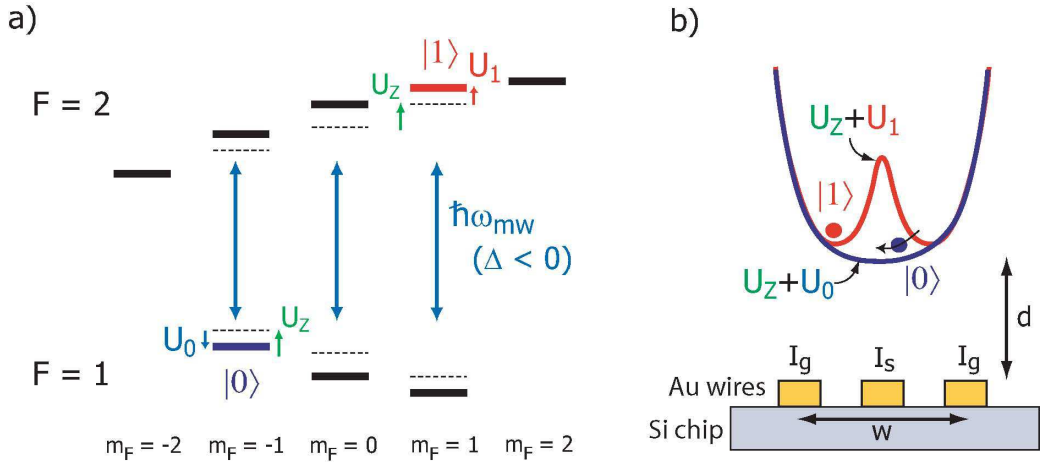}
\end{center}
\caption{\label{fig:MWpot} State-dependent microwave potentials for the qubit states. (a) Energy-level diagram of the hyperfine structure of the $^{87}$Rb ground state in a combined static magnetic and microwave field. $U_Z$ indicates the energy shift due to the static Zeeman effect, which is identical for $|0\rangle$ and $|1\rangle$. The magnetic field of the microwave couples the levels of $F=1$ to the levels of $F=2$, giving rise to energy shifts $U_1$ ($U_0$) for state $|1\rangle$ ($|0\rangle$), here shown for pure $\pi$ polarization and $\Delta < 0$ (red detuning). This shift has opposite sign for $|0\rangle$ and $|1\rangle$. (b) Chip layout and state-dependent double well potential for a collisional phase gate on the atom chip. The three gold conductors form a coplanar waveguide of width $w$ for the microwave. $I_s$ ($I_g$) are the currents on the signal (ground) wires. The wires carry both stationary and microwave currents, see text. In combination, these currents create the potential $U_Z+U_1$ for state $|1\rangle$ and $U_Z+U_0$ for state $|0\rangle$ at a distance $d$ from the chip surface.}
\end{figure}

In Fig.~\ref{fig:MWpot}a, this situation is shown for a $^{87}$Rb atom subject to a static magnetic field $\mathbf{B}_0(\mathbf{r})$, which defines the quantization axis, in combination with a microwave magnetic field $\mathbf{B}_\mathrm{mw}(\mathbf{r})\cos(\omega_\mathrm{mw} t)$. The static field gives rise to the static Zeeman potential $U_Z(\mathbf{r}) = \mu_B g_F m_F |\mathbf{B}_0(\mathbf{r})|$, which is identical for the qubit states $|0\rangle$ and $|1\rangle$, since for both states $g_F m_F = 1/2$. For simplicity, we assume that $\mathbf{B}_\mathrm{mw}(\mathbf{r})$ is oriented parallel to the static field $\mathbf{B}_0(\mathbf{r})$, corresponding to pure $\pi$ polarization of the microwave. The microwave field thus couples the transitions $|0\rangle \leftrightarrow |F=2,m_F=-1\rangle$ and $|F=1,m_F=1\rangle \leftrightarrow |1\rangle$ with identical resonant Rabi frequencies $\Omega_R(\mathbf{r})=\sqrt{3/4}\, \mu_B |\mathbf{B}_\mathrm{mw}(\mathbf{r})| / \hbar$. In the limit of large detuning $\hbar |\Delta| \gg \hbar \Omega_R, U_Z$, the coupling leads to microwave potentials given by
\[ U_1(\mathbf{r}) = -\frac{\hbar |\Omega_R(\mathbf{r})|^2}{4 \Delta} \quad \textrm{and} \quad U_0(\mathbf{r}) = \frac{\hbar |\Omega_R(\mathbf{r})|^2}{4 \Delta} \]
for $|1\rangle$ and $|0\rangle$, respectively. Since the qubit state $|0\rangle$ belongs to $F=1$ while $|1\rangle$ belongs to $F=2$, the microwave potential has opposite sign for the two states, giving rise to the desired state-dependence of the potential.

In a combined static magnetic and microwave trap, in general both $\mathbf{B}_0(\mathbf{r})$ and $\mathbf{B}_\mathrm{mw}(\mathbf{r})$ vary with position. This leads to a position-dependent microwave coupling with in general all polarization components present. If $\hbar |\Delta| \gg \hbar \Omega_R, U_Z$, the energy shifts due to the microwave coupling can be evaluated for each transition seperately. The overall magnetic microwave potential for the level $|F,m_F\rangle$ equals the sum of the energy shifts due to the individual transitions connecting to this level. The Zeeman splitting due to the static field (a few MHz) prevents two-photon transitions between sublevels $m_F$ belonging to the same $F$ quantum number.

A trap for neutral atoms based on microwave potentials has been proposed in \cite{Agosta89} and experimentally demonstrated in \cite{Spreeuw94}. This trap employs microwave radiation in the far field of the source. Unlike in the case of optical radiation, which can be tightly focussed due to its short wavelength, the centimeter wavelength $\lambda_\textrm{mw}$ of microwave radiation poses severe limitations on far-field traps: field gradients are very weak \cite{Spreeuw94} and structuring the potential on the micrometer scale is impossible.

On atom chips, there is a natural solution to this problem \cite{Treutlein04}. The atoms are trapped at distances $d\ll \lambda_\textrm{mw}$ from the chip surface.  Thus, they can be manipulated with microwave near fields, generated by microwave signals in on-chip transmission lines \cite{Collin01}. In the near field of the source currents and voltages, the microwave fields have the same position dependence as the static fields created by equivalent stationary sources. The maximum field gradients depend on the size of the transmission line conductors and on the distance $d$, not on $\lambda_\textrm{mw}$. In this way, state-dependent microwave potentials varying on the micrometer scale can be realized. In combination with state-independent static magnetic microtraps, the complex potential geometries required for QIP can be realized.

The state-dependent double well potential needed for the phase gate proposed in \cite{Calarco00} can be created with a chip layout as shown in Fig.~\ref{fig:MWpot}b. The three wires form a coplanar waveguide for the microwave. They carry both microwave and stationary currents, $I_s = I_c + I_\mathrm{mw} \cos(\omega_\mathrm{mw} t)$ and $I_g = I_o - (I_\mathrm{mw}/2) \cos(\omega_\mathrm{mw} t)$. The stationary currents $I_c$ and $I_o$ flow in opposite directions and create a static magnetic double well potential at a distance $d$ from the chip surface, as discussed in \cite{Reichel02}. We assume that the atoms are tightly confined in the transverse dimensions by a static magnetic potential created by additional wires not shown in the figure. The microwave currents create a microwave potential which is used to selectively remove the barrier of the double well for state $|0\rangle$, while increasing the barrier height for state $|1\rangle$ (Fig.~\ref{fig:MWpot}b). Note that for $\Delta<0$, as in the figure, the labeling of the states $|1\rangle$ and $|0\rangle$ is interchanged compared to \cite{Calarco00}.

To give a specific example, we consider atoms in a static-field trap at $d=1.8\,\mu$m from a microwave guiding structure of size $w=d$ carrying a microwave signal of amplitude $I_\mathrm{mw}=15$\,mA. A simulation of the microwave field yields a coupling with $\Omega_R/2\pi \sim 3.3$\,MHz at the position of the static double well barrier, taking into account the magnetic microwave field of the signal wire and both ground wires. For $\Delta = 10\, \Omega_R$, the change in the static magnetic moment of the qubit states due to the coupling is of the order of $10^{-3}$, such that both states still experience approximately the same static magnetic potentials. The microwave, on the other hand, leads to a differential energy shift of $U_1-U_0 = h \cdot 160$\,kHz, sufficiently large to remove the barrier for state $|0\rangle$. A detailed simulation for a realistic atom chip design shows that an improved version of the quantum phase gate of \cite{Calarco00} can be implemented with our robust qubit state pair using microwave potentials on the atom chip. We find an overall gate fidelity of $F=0.996$ at a gate operation time of $\tau_g=1.1$\,ms  \cite{Treutlein06tbp}, compatible with the requirements for fault-tolerant quantum computation.

\subsection{Qubit readout in microtraps}

The QIP schemes considered here use single atoms as qubit carriers,
and thus the final readout requires single-atom detectivity. Again,
the ability of atom chips to independently transport the individual
qubit atoms is a key advantage: atoms can be brought close together
for interaction, but spaced far apart and even transported to a remote
detector for readout. This removes the optical resolution limitation
that is still an unsolved problem in optical lattices. Thus, the basic
requirement on an atom chip qubit detector is single-atom detectivity
and compatibility with the presence of the chip. In our experiments,
we have focused on optical detectors, where fast progress could be
achieved. As an additional feature beyond single-atom detectivity, we
have concentrated on detectors that will ultimately allow quantum
non-demolition (QND) measurement of the number of atoms. A QND
trapped-atom detector would only perturb the phase of the atomic
state, but not, in particular, its vibrational energy in the trap.
Therefore, such a detector could also be used in qubit preparation,
for example in a ``feedback loop'' that prepares a single-atom state
from a larger initial BEC, by combining it with a switchable loss
mechanism.

To detect an atom optically, either absorption or dispersion can be
used. The collection of fluorescence light from a single trapped atom
is possible and has recently enabled remarkable experiments
\cite{Volz06,Beugnon06}. However, the recoil from the spontaneously
emitted fluorescence photons causes heating, ruling out the
possibility of a QND measurement. It might seem that single-pass
dispersive detection would offer a straightforward solution: the atom
trap would be positioned in one arm of an interferometer, operating at
a wavelength that is detuned far away from the atomic transitions.
However, to reach the high sensitivity required for single-atom
detection, a large number of photons must be sent through the
interferometer, and it turns out that even this type of detection
inevitably leads to spontaneous emission \cite{Lye03,Long03}. The
situation changes when an optical cavity is used to enhance the
interaction of the atom with the optical field. In this case,
single-atom detection with high signal-to-noise ratio is possible with
less than one spontaneous emission on average, and improves with high
cavity finesse $\mathcal F$ and small mode cross-section $w^2$.

This situation is adequately analyzed in the framework of cavity QED (cQED)
\cite{Kimble98}. The fundamental cavity QED parameters are the
coherent atom-photon coupling rate $g_0$, the cavity damping rate
$\kappa$ and the linewidth of the atomic transition $\gamma$.  For
single-atom detection, these parameters do not enter independently,
but in the combination $C=g_0^2/2\kappa\gamma$ called the
cooperativity parameter. The onset of the QND regime corresponds to
$C>1$.  Note that this condition is not identical with the strong
coupling regime of cQED, $g_0>\kappa,\gamma$. Indeed, QND detection is
possible even in the regime of weak coupling.

To translate the cooperativity criterion $C>1$ into requirements on
the cavity, it is instructive to analyze how $g_0$ and $\kappa$ relate
to the design parameters of the cavity. For a symmetric Fabry-Perot
(FP) cavity, these are the mirror radius of curvature $R$, the effective
cavity length $d$, and the cavity finesse $\mathcal F\approx
\pi/(T+\ell)$, where $T$ and $\ell$ are the transmission and losses of
a single mirror. One finds
\begin{eqnarray}
  \kappa &\propto& \mathcal F^{-1} d^{-1}\\
  g_0   &\propto& d^{-3/4}R^{-1/4}\\
  C      &\propto& (LR)^{-1/2}\\
\end{eqnarray}
$C$ can be alternatively expressed as
\begin{equation}
  C=\frac{3\lambda^2\mathcal F}{\pi^3 w^2}
\end{equation}
where $w$ is the mode waist diameter (and we assume that the atoms are
placed in this waist). This latter relation makes it intuitively clear
why $C$ is the relevant parameter for single-atom detection
efficiency: it is proportional to $\mathcal F$ (number of round-trips
of a cavity photon), and inversely proportional to the mode waist
diameter (which is to be compared to the atomic scattering
cross-section $\sigma$).  These relations hold within the stability
range of the cavity, and as long as the mode diameter on the mirrors
is small compared to the mirror diameter, so that clipping loss can be
neglected.

For the extremely short, small-volume cavities that we consider here,
$\gamma$ is always much smaller than $\kappa$ and $g_0$. Therefore, if
the goal is to enter as far as possible into the strong-coupling
regime, the cavity should optimize $g_0/\kappa$, i.e., increase the
mirror distance $d$ towards the limit of the stability range.  Indeed,
for a given mirror curvature, $\kappa$ drops as $\kappa\propto
d^{-1}$, whereas $g_0$ only decreases as $g_0\propto d^{-3/4}$, as
long as $d\ll R$: the ratio $g_0/\kappa$ increases with growing $d$
despite the decrease in the absolute value of $g_0$.  By contrast, a
cavity for single-atom detection should be designed to optimize the
cooperativity $C$.  According to the above proportionalities, this
means that it should have a short length and small radius of
curvature. A high finesse is obviously beneficial in both cases.

\subsubsection{Stable fiber Fabry-Perot cavities}
\begin{figure}[htb]
\begin{center}
\includegraphics[scale=0.6]{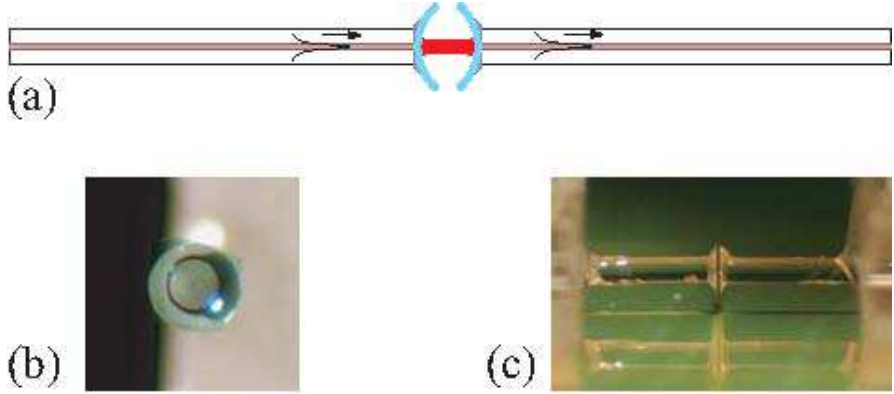}
\caption{(a) Concept of the stable FFP cavity. The
  basic building block is an optical fiber functionalized with a
  concave dielectric mirror. Two such fibers, brought sufficiently
  close to each other, result in a stable Fabry-Perot cavity which can
  be interrogated remotely, either in transmission or in reflection,
  through the two fibers (b) A
  single-mode optical fiber, total diameter 125 $\mu$m, processed with
  a concave mirror. The mirror has radius 1000 $\mu$m with a stopband
  centered at 780 nm.  (c) A complete FFP cavity, realizing the
  configuration (a), mounted on an atom chip used in the detection of
  cold atoms (Fig.~\ref{fig:FFPtrans}).}
\label{fig:FFPconcept}
\end{center}
\end{figure}

The ``gold standard'' for cQED cavities is still being set by
macroscopic FP cavities with superpolished, concave
mirrors. These mirrors have relatively large radii of curvature
($R=20\,$cm is typical) and achieve record finesse values of
$\mathcal{F}> 2\times 10^6$ \cite{Rempe92}. However, these cavities
are not compatible with a chip-based microtrap. The trap-surface
distance is $\lesssim 250\,\mu$m, whereas the diameter of existing
superpolished FP mirrors is at least $\sim 1\,$mm, so that it would be
extremely difficult to place the optical axis sufficiently close to
the substrate surface and still maintain the tight mirror spacing
required for high $C$. We have developed stable, fiber-based
Fabry-Perot resonators (FFPs) \cite{Long03} that avoid this problem.
They employ concave dielectric mirror coatings with small radius of
curvature, realized on the fiber tip. A stable cavity is constructed
from two closely spaced fiber tips placed face-to-face
(figure~\ref{fig:FFPconcept}(a)).  Thus, as an important difference to other
microcavities such as microtoroid resonators (see for example
\cite{Spillane05}), the cavity mode is located in free space between
the fibers, thus avoiding the extremely restrictive positioning
requirements imposed by evanescent-field coupling.

\subsubsection{FFP cavity fabrication and performance}
We have fabricated stable, miniature FFP cavities using two different
methods. Method 1 uses a commercially available lift-off coating
\cite{Steinmetz06tbp}.  The coating is produced on a convex template
(we use a commercial ball lens), and then glued onto the fiber tip.
After curing the transparent epoxy glue with UV light, the coating
sticks to the fiber and a small force is enough to detach it from the
ball lens template.  The result is a fiber functionalized with a
highly reflecting concave mirror, as shown in
Fig.~\ref{fig:FFPconcept}(b). A complete FFP cavity is shown in
Fig.~\ref{fig:FFPconcept}(c). This method reproducibly leads to cavity
finesse values $\mathcal F>1000$ with modest experimental effort. We
have used a cavity of this type to detect magnetically guided and
trapped atoms, as described below.  Method 2 employs laser surface
processing to produce a low-roughness concave depression on the fiber
tip, followed by multilayer coating using the ion beam sputtering
technique. With this technique, we obtain finesse values $\mathcal
F\sim 35000$ \cite{Hunger06tbp}.

In both cases, because of the small fiber diameter (125\,$\mu$m),
very short cavities ($<10 \lambda/2$) can be realized even when radii
of curvature $R\le1\,$mm are used, still leaving a sufficiently large
gap to introduce cold atoms. Let us consider the concrete example of a
cavity that we have fabricated using method 1. The mirror curvature is
$R=1\,$mm and the distance $d=27\,\mu$m, leading to a mode volume
$V_m=600\,\mu\textrm{m}^3$, to be compared to
$V_m=1680\,\mu\textrm{m}^3$ for the smallest-volume macroscopic FP
cavity that has been used with atoms \cite{Hood00}.  In terms of
cavity QED parameters, the small mode volume results in an
exceptionally high coherent atom-photon coupling rate,
$g_0/2\pi=180\,$MHz (calculated for the Rb D2 line at
$\lambda=780\,$nm). Therefore, in spite of a comparatively high
cavity damping rate $\kappa/2\pi=2.65\,$GHz, which results from the
moderate finesse of the transfer coating and short cavity length, the
cavity reaches a single-atom cooperativity parameter greater than
unity, $C=g_0^2/2\kappa\gamma=2.1$, (for the Rb D2 line, $\gamma/2\pi = 3.0$\,MHz) signaling the onset of quantum
effects such as enhanced spontaneous emission into the cavity mode
\cite{Kimble98}  and a significant modification of cavity transmission by
the presence of a single atom.

Below we present an experiment in which we use the two-fiber cavity to
detect an extremely small number of cold atoms magnetically trapped in
the cavity using an atom chip. What is still missing is an improved
absolute calibration of these results in order to determine whether
they already realize, or only come close to single-atom detectivity.
In any case, considering that the ``method 2'' cavities are now
available and improve finesse by a factor 30, it seems clear that the
problem of qubit detection can be solved using our FFP technology.
Beyond QIP, we believe that this cavity type is also attractive
for experiments exploiting the strong optical dipoles of
semiconductor quantum dots, semiconductor nanocrystals and molecules,
and for channel separation in telecommunication.

\subsection{On-chip atom detection with a FFP cavity}

\begin{figure}[h]
\begin{center}
\includegraphics[scale=0.35]{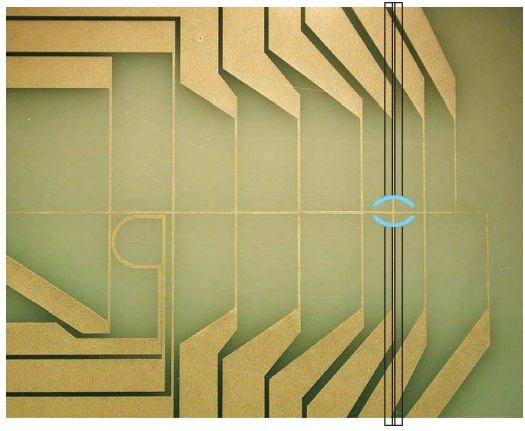}
\caption{The chip used for detecting magnetically trapped atoms by
  their interaction with a single optical mode in a fiber Fabry-Perot
  (FFP) cavity. The position of the cavity is indicated in exaggerated
  size for better visibility. Atoms are first trapped in the
  structures on the left side of the picture, then transported towards the cavity and evaporatively
  cooled in a multistep procedure and finally positioned in the
  resonator mode, where they are detected by the change in cavity
  transmission.}
\label{fig:FFPchip}
\end{center}
\end{figure}

We have detected magnetically trapped atoms with an FFP
cavity on an atom chip \cite{Steinmetz06tbp}. The atoms are trapped on
the chip and evaporatively cooled as in our previous experiments
\cite{Treutlein04}, but on a chip which incorporates the FFP resonator
fabricated according to method 1 described above
(Fig.~\ref{fig:FFPchip}). Trapped and guided atoms could be
reproducibly detected in a great variety of experimental parameters
and procedures. The cavity transmission signal allowed detection with
good signal-to-noise ratio even when the atom number was far too small
to be visible by our absorption-imaging camera system. A typical
temperature of the atom cloud in the resonator was around $1\,\mu$K,
with typical longitudinal and transverse trap frequencies around
100\,Hz and 1\,kHz, respectively. Clouds containing extremely few
atoms were prepared using the RF knife, by applying repeated, rapid
radiofrequency scans across the ``trap bottom'' frequency.

\begin{figure}[h]
\begin{center}
\includegraphics[scale=0.6]{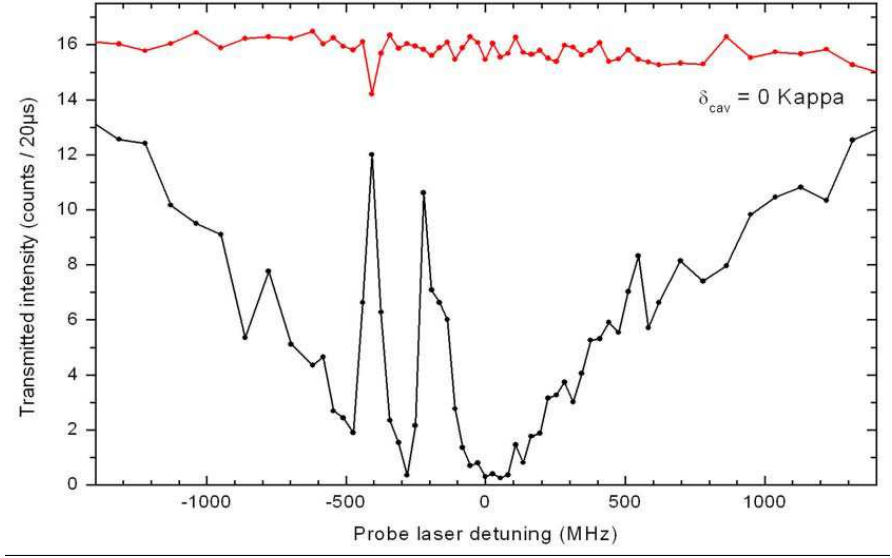}
\caption{Cavity tranmission spectrum without atoms (upper, red curve)
  and with atoms magnetically trapped in the on-chip FFP cavity
  (lower, black curve). The atom-cavity detuning is fixed at
  $\delta_\textrm{cav}=0$. Each point corresponds to a complete
  experimental sequence of preparation, evaporative cooling,
  positioning and detection, for the atom-laser detuning indicated on
  the abscissa. Lines are to guide the eye. Zero probe laser detuning
  corresponds to the $F=2\rightarrow F'=3$ transition within the D2
  multiplett.}
\label{fig:FFPtrans}
\end{center}
\end{figure}

Figure~\ref{fig:FFPtrans} shows a cavity transmission spectrum
recorded by scanning the probe laser across the D2 atomic transition
for a fixed atom-cavity detuning of $\delta_{\textrm{cav}}=0$. Each
point in the spectrum corresponds to a complete experimental sequence
of preparation, evaporative cooling, positioning and detection.  The
atoms are initially trapped in the $|F=2,m=2\rangle$ ground state. The three transmission minima correspond to transitions from this state to the
$F=1,2,3$ sublevels of the $5p_{3/2}$ excited state.  We have
recorded such spectra for various $\delta_{\textrm{cav}}$ and for
different mean atom numbers.

The parameter of interest is the minimum number $N_{\min}$ of atoms
that must interact with the cavity mode in order to produce a
detection signal with a good signal-to-noise ratio, say, 4. The actual
number of detected atoms is more difficult to extract from the
measurements than with a macroscopic resonator. This is mainly due to
large error bars on the measured on-resonance cavity transmission,
which in turn are caused by the fact that incoupling mirror and fiber
cannot be separated. For the resonator used in the experiment
described here, the on-resonance transmission is in the permille
range. This is due to an excessive number of layers in the dielectric
mirror stack, applied by the coating manufacturer in an attempt to
maximise the reflectivity. This problem no longer occurs with the
cavities fabricated later according to method 2. In the experiments
described here, however, the low transmission means that for every
detected photon, roughly $10^3$ photons have interacted with the atom
without contributing to the signal. They do, however, contribute to
heating due to spontaneous-emission, and we therefore expect the
detectivity in this experiment to be limited by this spontaneous
heating.  Nevertheless, we expect $N_{\min}$ to be close to or below 1.

From atom number measurement by absorption imaging, we can infer an
upper limit of $N_{\min}$ which is of the order of 50 atoms. A much
more precise value of $N_{\min}$ can be obtained from spectra such as
in Fig.~\ref{fig:FFPtrans}. These spectra were obtained with an FFP
cavity of relatively low finesse $\mathcal F\sim 260$. This
corresponds to a weak-coupling regime in which the atom-cavity
interaction can be understood semiclassically. The spectra depend very
strongly on the mean atom number in the cavity. We are now using a
semiclassical model to fit the spectra, which will allow us to
determine the actual number of intracavity atoms with good precision
without the need to know the absolute cavity transmission. In this
way, we will be able to determine the detectivity from the
experimental results.

These results demonstrate the suitability of FFP resonators for qubit
readout on atom chips. The combination of this new cavity type with
atom chips will enable new applications beyond atom detection. The
laser-machined resonators, (method 2 described above), which we are
now integrating into an atom chip experiment, reach a finesse
$\mathcal F\sim 35000$, combined with an exceptionally small mode
volume. For these resonators, with a mirror spacing of $d=25\,\mu$m,
one obtains $g_0\sim 2\pi\times 400\,$MHz and $g_0/\kappa\sim 4$,
entering the strong-coupling regime of cavity QED.  But even in the
regime of weak coupling, trapping an ultracold atom cloud in an
optical cavity of high cooperativity, as demonstrated here, is a new
experimental option which can radically simplify the implementation of
high-fidelity atom-photon interfaces, for example in quantum
communication \cite{Duan01}.

\subsection{Single atom preparation}

With the advent of single-atom detectors on atom chips, it becomes
possible to address the problem of deterministic single-atom
preparation. For the QIP schemes considered here, each qubit is a
single atom \emph{ in the ground state of a magnetic potential}. A
first, simplistic approach is to start from a BEC and induce losses to
reduce the atom number to an average value of 1. With a QND detector,
the actual number can be measured, and further reduced if necessary,
with negligible excitation and loss. Nevertheless, this
``trial-and-error'' method becomes impractical for large numbers of
qubits.  Proposals for deterministic single atom preparation have been
put forward in \cite{Diener02,Mohring05}. The key element in these
methods is a tightly confining potential, in which states with 1, 2
etc. atoms are energetically resolved due to the collisional
interaction. A BEC serves as a reservoir from which single atoms can
be repeatedly extracted in a deterministic way. Atom chips appear
ideally suited to implement this idea, and we expect it to be
experimentally realized within the next two years.

\section{Conclusion}

The fast experimental progress made with atoms in optical lattices and magnetic microtraps  underlines the great potential of ultracold quantum gases for applications in QIP. In the experiments with optical lattices described here, a massively parallel quantum gate array was demonstrated for the first time \cite{Mandel03b}, which allows the creation of a highly entangled many-body cluster state. In the future, it is important to explore quantum computing schemes which rely only on single-atom operations and measurements on the entangled many-body state. New theoretical developments show that even without the possibility of performing single-atom manipulations in the optical lattice, a quantum computer based on the controlled collisions demonstrated here could simulate a large class of complex Hamiltonians with translational invariance, which play an important role in condensed-matter physics.

A general quantum computer, however, requires the possibility to perform single-atom operations and measurements. The fiber Fabry-Perot resonators described here are an ideal system for achieving this goal. The detection of very small atom numbers was demonstrated in our experiments with a FFP resonator integrated on the atom chip \cite{Steinmetz06tbp}. It seems clear that the problem of single qubit detection can be solved in the nearest future with the technical improvements of this detector which have been recently implemented \cite{Hunger06tbp}. We have furthermore shown that using a qubit state pair which is robust against magnetic-field fluctuations, coherence lifetimes exceeding one second can be achieved on an atom chip with atoms at distances down to a few microns from the chip surface \cite{Treutlein04}. Based on these developments, the main experimental challenges for the future are the reproducible preparation of single-atom states and the implementation of a quantum phase gate using microwave potentials on the atom chip. The theoretical fidelity of such a gate is $0.996$ \cite{Treutlein06tbp}, compatible with the requirements for fault-tolerant quantum computation.

The success of these future experiments will determine whether QIP with neutral atoms is an advantageous alternative to other systems such as trapped ions, and allows the experimental investigation of even more complex problems such as quantum error correction.

   This work has been funded by the Bavarian State Government
  (Kompetenznetzwerk Quanteninformation ``A8''). Y.C. gratefully
  acknowledges support from the ``CONQUEST'' network
  (MRTN-CT-2003-505089).

%\bibliographystyle{../bib/jrprsty}
%\bibliography{../bib/lascool}
%\end{document}

\end{document}